\documentstyle [12pt]{article}
\begin{document}
\title{Real decoupling ghosts quantization of CGHS model for
two-dimensional\\black holes.\footnote{Published Phys.Rev.D 51 n 4; 1995}}
\author{S. Liberati\\
Dipartimento di Fisica,\\
Universita' di Roma ``La Sapienza"\\
and Istituto Nazionale di Fisica Nucleare, Sezione di Roma I\\
00185 Roma, Italy}
\date{1994}
\maketitle
\begin{abstract}
A complete RST quantization of a CGHS model plus Strominger term is carried
out.
In so doing a conformal invariant theory with $\kappa=\frac{N}{12}$ is found,
that is, without ghosts contribution. The physical consequences of the model
are analysed and positive definite Hawking radiation is found.
\end{abstract}
\section{Introduction}
In the past two years a great advance in the studies of two-dimensional models
of black hole has taken place. One of the most interesting approaches to this
problem is certainly the model proposed by Callan, Giddings, Harvey, Strominger
\cite{CGHS} in 1992. In that work the authors have introduced a two-dimensional
action involving gravity coupled to a dilaton field and N conformal matter
fields:
\begin{equation}
S_{cl}=\frac{1}{2\pi}\int d^{2}x \; \sqrt{-g}[e^{-2\phi}(R+4(\nabla\phi)^{2}
+4 \lambda^{2})-\frac{1}{2}\sum_{i=1}^{N}(\nabla f_{i})^{2}]
\end{equation} In the conformal gauge $g_{++}=g_{--}=0,g_{+-}=-\frac{1}{2}
e^{2\rho}$ this becomes:
\begin{equation}
S_{cl}=\frac{1}{2\pi}\int d^{2}x \; [e^{-2\phi}(2\partial_{+}\partial_{-}\rho-4
\partial_{+}\phi\partial_{-}\phi+\lambda^{2}e^{2\rho})+\frac{1}{2}
\sum_{i=1}^{N} \partial_{+}f_{i}\partial_{-}f_{i}]
\end{equation}
Those authors have shown that the model admits solutions
describing the formation of an evaporating black hole and that it is possible
to give a semiclassical treatment of Hawking radiation. There was a great
interest around this model, expecially about how to quantize it to arrive to a
fully quantum treatment of Hawking radiation (albeit in two dimensions). There
are two kinds of approach to the quantization problem, the first of which was
proposed by Bilal, Callan and de Alvis \cite{BC,A}. They quantized the
classical action (2) by generalysing the form of the potential to obtain a
conformal field theory, and so one loop finite. One of the problems of this
approach is the lack of the linear dilaton vacuum (LDV) as a solution (the
corrisponding solution they find is only an asymptotic one). They also found
an eternal radiating black hole and consequentely the appearance of arbitrary
large negative energies.

The other approach was proposed by Russo, Susskind and Thorlacius \cite{RST}.
They modified the classical action (2) to obtain a theory with the linear
dilaton vacuum as a solution. In order to do this, they added at one loop:
\begin{equation}
S_{RST}=-\frac{\kappa}{\pi}\int d^{2}x \; \phi\partial_{+}\partial_{-}\rho
\end{equation}
This is a local covariant term that allows the preservation, also at one loop,
of the classically conserved current:
\begin{equation}
J_{\pm}=\partial_{\pm}(\rho-\phi)\ \ with \ \ \partial_{+}\partial_{-}(\rho-
\phi)=0
\end{equation}
Its existence allows in their model the persistence at one loop level of the
LDV solution and the linking of the evaporating black hole solution to a
"shifted" LDV, terminating, in a physical reasonable way, the flux of Hawking
radiation in finite time. The price of this is a violation of the "cosmic
censorship" hypothesis, or at least the radical redefinition of it
\cite{C.Cens}.

Apart from the difficulties explained above, the two models share the same
problem: to obtain a conformal field theory they have to fix the value of the
coefficient of the matter anomaly term $\kappa$ to be:
\begin{equation}
\kappa=\frac{N-24}{12}
\end{equation}
The shift in $\kappa$ as compared to the semiclassical value $\kappa=
\frac{N}{12}$ found by CGHS, is a consequence of the missing term in these
theories which would be responsible for decoupling the contribution of the
reparametrization ghosts. The shift in the value of $\kappa$ induces spurious
mode in Hawking radiation (a negative unphysical contribution) and causes a
critical dependence on the number of matter fields ($N>$or$<24$).
Strominger \cite{Str} has proposed a method to obtain a decoupling of ghost
fields by adding a term to the classical action. He proposed to improve the
action with:
\begin{equation}
S_{Ghost}=\frac{1}{\pi}\int d^{2}x \; [2\partial_{+}(\rho-\phi)
\partial_{-}(\rho-\phi)]
\end{equation}
His basic point was the observation that there is no fundamental reason for
choosing the same metric $g_{ij}$ to define the measure of the path integral
of the graviton-dilaton-ghost as that used for the matter fields, since the
difference amounts to the addiction of a local function of Weyl factor. So he
used a rescaled metric $\hat{g}_{ij}=g_{ij}e^{-2\phi}$ to define the measure
of the graviton-dilaton-ghost system (the Faddev-Popov ghost determinant).
This metric appears to be flat for all classical solutions of the original
CGHS theory. The consequence is that black holes do not radiate ghosts to
leading order in this theory. In conformal gauge this idea is implemented by
building the graviton-dilaton-ghost Polyakov term out of $\rho-\phi$ (the
action (6)) while building the matter anomaly term out of $\rho$ (the usual
matter generated Polyakov term yet proposed by CGHS):
\begin{equation}
S_{anom}=-\frac{\kappa}{\pi}\int d^{2}x \; \partial_{+}\rho \partial_{-}\rho
\end{equation}

Unfortunately it appears that in quantizing this theory "\`{a} la ABC" (Bilal,
Callan, de Alvis \cite{BC,A}) the value of $\kappa$ has to be fixed, to obtain
a
conformal invariant field theory, again as $\kappa=\frac{N-24}{12}$ showing
that the decoupling of the ghost is not complete \cite{BC}.

Bilal \cite{B} has explained this defect of Strominger model by the fact that
the equations of motion derived from $S_{cl}+S_{anom}+S_{Ghost}$ differ, for
$N=0$, from those derived from $S_{cl}$ only by terms $\approx \partial_{+
}\partial_{-}(\rho-\phi)$ that vanish by the presence of the conserved current
(4). This is not true for $N\neq0$ (in ABC quantization). So only in the
absence of the matter fields $S_{Ghost}$ does not contribute to the solution
of those equations.

Here I propose a RST quantization of a generalized Strominger model; in so
doing the $\partial_{+}\partial_{-}(\rho-\phi)=0$ condition is automatically
satisfied also at one loop level and result is a really ghost decoupled
conformal field theory. In the next section the new model will be described
in detail. In sections III and IV its behaviour will be studied and in section
V predictions about Hawking radiation will be made.

\section{The Model}
The starting point is the usual CGHS action (2). At one loop level this
becomes in the new model:
\begin{equation}
S=S_{cl}+S_{anom}+S_{RST}+S_{Strom}
\end{equation}
where:
\begin{eqnarray*}
S_{cl}&=&\frac{1}{2\pi}\int d^{2}x \;
[e^{-2\phi}(2\partial_{+}\partial_{-}\rho-
4\partial_{+}\phi\partial_{-}\phi+\lambda^{2}e^{2\rho})+
\frac{1}{2} \sum_{i=1}^{N} \partial_{+}f_{i}\partial_{-}f_{i}] \\
S_{anom}&=&-\frac{\kappa}{\pi}\int d^{2}x\; \partial_{+}\rho \partial_{-}\rho\\
S_{RST}&=&-\frac{\kappa}{\pi}\int d^{2}x \; \phi\partial_{+}\partial_{-}\rho\\
S_{Strom}&=&\frac{1}{\pi}\int d^{2}x \; [2\partial_{+}(\rho-\phi)\partial_{-}
(\rho-\phi)]
\end{eqnarray*}
It is easy to recognise in $S_{anom}$ the usual anomalous Polyakov term of the
matter action, in $S_{RST}$ the covariant term modifying the kinetic part of
the classical action "\`{a} la RST" \cite{RST} (thanks to this term
$\partial_{+}\partial_{-}(\rho-\phi)=0$ also at one loop level), and in
$S_{Strom}$ the above mentioned Strominger term.
The costraint equations of the conformal gauge give:
\begin{eqnarray*}
T_{\pm\pm}\!&=&\![e^{-2\phi}+\frac{\kappa}{4}](4\partial_{\pm}\rho\partial_{\pm}
\phi-2\partial_{\pm}^{2}\phi)+\frac{1}{2}\sum_{i=1}^{N}\partial_{\pm}f_{i}
\partial_{\pm}f_{i}+\\
&-&\!\kappa(\partial_{\pm}\rho\partial_{\pm}\rho-
\partial_{\pm}^{2}\rho)+2[\partial_{\pm}(\rho-\phi)\partial_{\pm}(\rho-\phi)-
\partial_{\pm}^{2}(\rho-\phi)]+t_{\pm}=0
\end{eqnarray*}
that is:
\begin{eqnarray}
T_{\pm\pm}\!&=&\![e^{-2\phi}+\frac{\kappa}{4}-1]
(4\partial_{\pm}\rho\partial_{\pm}\phi-2\partial_{\pm}^{2}\phi)+
\frac{1}{2} \sum_{i=1}^{N} \partial_{\pm}f_{i} \partial_{\pm}f_{i}+ \nonumber
\\
&-&\!(\kappa-2)(\partial_{\pm}\rho\partial_{\pm}\rho-
\partial_{\pm}^{2}\rho)+2\partial_{\pm}\phi\partial_{\pm}\phi+t_{\pm}=0
\end{eqnarray}
It is important to underline that the form of the stress energy tensor given
in (9) can be determined only from the covariant expression of the action by
derivations as to the components of the metric.

The function $t_{\pm}$ appearing in (9), arises by the non locality of the
Polyakov term $\approx\int d^{2}x \; R (\partial^{\mu}\partial_{\mu})^{-1}R$
that generates the anomalous part; $t_{\pm}$ has to be determined by imposing
boundary condictions.

Following Bilal and Callan's method \cite{BC} we perform a redefinition of the
fields to obtain a Liouville-like theory. This brings to a new relation
between physical fields $\rho$, $\phi$ and "Liouville" ones $\Omega$,
$\chi$:

\begin{equation}
\left\{\begin{array}{rcl}
\chi \!&=&\! \rho \sqrt{\kappa-2}+\frac{\textstyle e^{-2\phi}}
{\textstyle \sqrt{\kappa-2}}-\frac{\textstyle \kappa-4}{\textstyle 2\sqrt
{\kappa-2}} \phi\\
&&\\
\Omega \!&=&\! \frac{\textstyle e^{-2\phi}}{\textstyle \sqrt{\kappa-2}}+
\frac{\textstyle \kappa}{\textstyle 2\sqrt{\kappa-2}} \phi\\
\end{array}
\right.
\end{equation}
\\
With this redefinition the action is transformed into the Liouville-like form:
\begin{equation}
S=\frac{1}{\pi}\int d^{2}x \; [-\partial_{+}\chi\partial_{-}\chi+
\partial_{+}\Omega \partial_{-}\Omega+
\lambda^{2}{\textstyle e}^{\frac{2}{\sqrt{\kappa-2}}
(\chi-\Omega)}+
\frac{1}{2}\sum_{i=1}^{N} \partial_{+}f_{i}\partial_{-}f_{i} \;]
\end{equation}

and the stress energy tensor (9) becomes:

\begin{equation}
T_{\pm}=-\partial_{\pm}\chi\partial_{\pm}\chi+\partial_{\pm}\Omega
\partial_{\pm}\Omega+\sqrt{\kappa-2}\;\partial_{\pm}^{2}\chi+
\frac{1}{2}\sum_{i=1}^{N} \partial_{\pm}f_{i}\partial_{\pm}f_{i}
\end{equation}

{}From the expression of stress energy tensor (12) one can easily determine the
value of $\kappa$ required to obtain a conformal field theory. We have:

\begin{eqnarray}
c=c_{\chi}+c_{\Omega}+c_{M}+c_{ghost}\!&=&\![1-12(\kappa-2)]+1+N-26=0 \nonumber
\\
\mbox{for} \: \kappa\!&=&\!\frac{N}{12}
\end{eqnarray}

This shows that the model is a conformal invariant field theory with $\kappa$
equal to the semiclasssical value found by CGHS without ghosts shift.
One can immediately see that also in this case the field redefinition (10) is
not "one to one" and in fact $\Omega$ is bounded from below for
$-\infty<\phi<\infty$.
Thus, in the present model too, there is a minimum for $\Omega'=0$ which
corresponds to
\begin{equation}
\phi=-\frac{1}{2}\ln \left( \frac{\kappa}{4} \right)=
-\frac{1}{2}\ln \left( \frac{N}{48} \right) \equiv\phi_{c}
\end{equation}

It is obvious that something singular happens for this value of the field
$\phi$. It is also important to note that this behaviour of $\Omega$ field,
is common to all models cited above (ABC, RST). According to RST approach, this
will bring us to fix a boundary condition that restricts the range of $\Omega$
from plus infinity to the critical value. Of course, as pointed out by Hawking
\cite{Haw}, in so doing we don't have a real Liouville theory but rather a
Liouville-like one, because the RST condition brings to an effectively non
linear theory.
The solutions of Liouville theory will only be a first approximation of the
real ones.

Let us now study the physical behaviour of the model to test if it presents,
as it is expected, results independent from the ghost contribution.

\section{Solutions}
The equations of motion of model (8) are:
\begin{equation}
\left\{\begin{array}{lrl}
\chi:& -2\partial_{+}\partial_{-}\chi \! &= \!
\frac{\textstyle 2\lambda^{2}{\textstyle e}^{\frac{2}
{\sqrt{\kappa-2}}
(\chi-\Omega)}}
{\textstyle \sqrt{\kappa-2}}\\
&&\\
\Omega:& +2\partial_{+}\partial_{-}\Omega \! &= \!
-\frac{\textstyle 2\lambda^{2}{\textstyle e}^{\frac{2}
{\sqrt{\kappa-2}}
(\chi-\Omega)}}
{\textstyle \sqrt{\kappa-2}}
\end{array}
\right.
\end{equation}
\noindent
These correspond to:
\begin{equation}
\left\{\begin{array}{rcl}
\partial_{+}\partial_{-}(\chi-\Omega)\! &=& \!0\\
&&\\
\partial_{+}\partial_{-}(\chi+\Omega)\! &=& \!
- \frac{\textstyle 2 \lambda^{2}{\textstyle e}^{\frac{2}{\sqrt{\kappa-2}}
(\chi-\Omega)}}
{\textstyle \sqrt{\kappa-2}}
\end{array}
\right.
\end{equation}
\noindent
The first equation of (16) is what one expects to find thanks to the RST term,
in fact now $\partial_{+}\partial_{-}(\rho-\phi)=0$ implies
$\partial_{+} \partial_{-}(\chi-\Omega)=0$ as it is easy to verify using fields
redefinition (10). As a result of this a gauge can be chosen in which
$\chi=\Omega$, $\rho=\phi$, the second equation thus reduces to:

\begin{equation}
\partial_{+}\partial_{-}\Omega=-\frac{\lambda^{2}}{\sqrt{\kappa-2}}
\end{equation}
\noindent
Following the preceding works cited above we look for solution of (17)
corresponding to flat static geometry. These are of the form:

\begin{equation}
\Omega=\chi=-\frac{\lambda^{2}x^{+}x^{-}}{\sqrt{\kappa-2}}+
\frac{P\kappa}{\sqrt{\kappa-2}}\ln(-\lambda^{2}x^{+}x^{-})+
\frac{M}{\lambda\sqrt{\kappa-2}}
\end{equation}
\noindent
Different values of P and M correspond to different solutions of the model:
1 \qquad For $P=-\frac{1}{4}$, $M=0$ one finds the usual LDV solution:

\begin{equation}
\Omega=\chi=-\frac{\lambda^{2}x^{+}x^{-}}{\sqrt{\kappa-2}}-
\frac{\kappa}{4\sqrt{\kappa-2}}\ln(-\lambda^{2}x^{+}x^{-})
\end{equation}
\noindent
that is:

\begin{equation}
e^{-2\phi}=e^{-2\rho}=-\lambda^{2} x^{+}x^{-}
\end{equation}
2 \qquad For $P=0$, $M\neq0$ one finds the usual quantum black hole in
thermal equilibrium with its environments:

\begin{equation}
\Omega=\chi=-\frac{\lambda^{2}x^{+}x^{-}}{\sqrt{\kappa-2}}+
\frac{M}{\lambda\sqrt{\kappa-2}}
\end{equation}
\noindent
The beaviour of solutions (21) is the same as that shown in the preceding
works \cite{BC,RST,B}.

\section{Evaporating black holes}

Following the work of RST \cite{RST}, let us consider a dynamical situation
where an incoming shock wave carries energy into a linear dilaton vacuum along
an infalling line $x^{+}=x_{0}^{+}$. Putting in the ++ costraints:

\begin{equation}
\frac{1}{2} \sum_{i=1}^{N} \partial_{+}f_{i}\partial_{-}f_{i}=
\frac{m}{\lambda x_{0}^{+}} \delta(x^{+}-x_{0}^{+})
\end{equation}
\noindent
we obtain:
\begin{equation}
\Omega=\chi=-\frac{\lambda^{2}x^{+}x^{-}}{\sqrt{\kappa-2}}-
\frac{\kappa}{4\sqrt{\kappa-2}}\ln(-\lambda^{2}x^{+}x^{-})-
\frac{m(x^{+}-x_{0}^{+})}{\lambda x_{0}^{+} \sqrt{\kappa-2}} \theta(x^{+}-
x_{0}^{+})
\end{equation}
\noindent
The solution (23) describes an evaporating black hole.
In $\Omega$ and $\chi$ the singularity is not evident, but if we work in
original fields $\phi$ and $\rho$ it is easy to see it. If we consider the
curvature scalar $R=8e^{-2 \rho} \partial_{+} \partial_{-} \rho$ we find that
it is singular for $\Omega'=0$ that is
$\phi=-\frac{1}{2} \ln (\frac{N}{48}) \equiv \phi_{c}$ mentioned before (14).
So the singularity lies on the curve $\phi=\phi_{c}$ where $\Omega$ takes the
value:

\begin{equation}
\Omega=\Omega_{c}=\frac{\kappa}{4\sqrt{\kappa-2}} \left[ 1-\ln \left(
\frac{\kappa}{4} \right) \right]
\end{equation}
\noindent
Introducing (24) in the dynamical solution (23) we find the same curve of the
singularity $(\overline{x}^{+},\overline{x}^{-})$ found by RST \cite{RST}:
\begin{equation}
1-\ln \left( \frac{\kappa}{4} \right)=
-\frac{4\lambda^{2}}{\kappa} \overline{x}^{+} \overline{x}^{-}-
\ln(-\lambda^{2} \overline{x}^{+} \overline{x}^{-})-
\frac{4m}{\lambda x^{+}_{0} \kappa} ( \overline{x}^{+}-x_{0}^{+})
\theta  ( \overline{x}^{+}-x_{0}^{+})
\end{equation}
\noindent
This curve asymptotically approaches the line:
\begin{equation}
x^{-}=-\frac{m}{\lambda^{3}x_{0}^{+}}
\end{equation}
\noindent
Timelike observers with $x^{-}>-\frac{m}{\lambda^{3}x_{0}^{+}}$ could not
escape from the singularity, so line (26) is an equivalent of the classical
global event horizon for them. The classical horizon is also a curve where the
dilaton field's gradient passed from spacelike to timelike. Our region where
$\nabla \phi$ is timelike corresponds to a trapped region in higher
dimensional theory. So in two dimensions we define an apparent horizon as the
line where $\nabla \phi=0$. So doing the apparent horizon coincides with the
event horizon of a static solution.
\noindent
Considering the curve $(\hat{x}^{+},\hat{x}^{-})$ where $\nabla \phi=0$ we
find, as RST:
$$\partial_{+} \phi=0 \Longrightarrow \partial_{+} \Omega=0$$
\noindent
for
\begin{eqnarray*}
0\!&=&\!-\frac{\lambda^{2} x^{-}}{\sqrt{\kappa-2}}-\frac{\kappa}
{4\sqrt{\kappa-2}}
\frac{-\lambda^{2} x^{-}}{-\lambda^{2} x^{+} x^{-}}+ \nonumber \\
&-&\overbrace{\frac{m(x^{+}-x_{0}^{+})}{\lambda x_{0}^{+} \sqrt{\kappa-2}}
\delta (x^{+}-x_{0}^{+})}^{=0}-\frac {m \theta(x^{+}-x_{0}^{+})}{\lambda
x_{0}^{+} \sqrt{\kappa-2}}\\
\end{eqnarray*}
\noindent
so
\begin{eqnarray*}
-\frac {\kappa}{4x^{+}\sqrt{\kappa-2}}
=\frac {\lambda^{2}x^{-}}{\sqrt{\kappa-2}}+
\frac {m \theta(x^{+}-x_{0}^{+})}{\lambda x_{0}^{+} \sqrt{\kappa-2}}
\end{eqnarray*}
\noindent
that is:
\begin{equation}
\hat{x}^{+}=-\frac{\kappa}{4\lambda^{2}} \frac{1}{\hat{x}^{-}+
\frac{m}{\lambda^{3} x_{0}^{+}}} \:\:\:\:\:
\mbox{for}\:\:\:\: x^{+}>x_{0}^{+}
\end{equation}

Let us now study the behaviour of the horizon. From (27) we find:
\begin{equation}
\hat{x}^{-}=-\frac{\kappa}{4\lambda^{2}x^{+}}-\frac{m}{\lambda^{3} x_{0}^{+}}
\end{equation}
\noindent
so
\begin{equation}
\frac{d \hat{x}^{-}}{d x^{+}}=\frac{\kappa}{4\lambda^{2} {x^{+}}^{2}}=
\frac{N}{48\lambda^{2} {x^{+}}^{2}}>0
\end{equation}
\noindent
This is an important result because it tells us that the apparent horizon
recedes also in this model at a rate proportional to $\kappa$ alone.
What is new now is that $\kappa$ is not affected by the ghost's contribution ($
\kappa=\frac{N-24}{12}$ in ABC and RST model). The rate of contraction of the
horizon is directly related to Hawking radiation so we expect to find a ghost
independent energy flux as we shall prove in section V.

As shown by RST \cite{RST}, the two curves (25)(27) determined above meet for
a finite value of $x^{+}$ and from this point on evaporating black hole
solution can be matched with a linear dilaton one shifted by a little amount
of negative energy. This is  equivalent, in the $--$ costraints, to a little
shock wave of matter with negative energy propagating out: the RST
"Thunderpop". A brief analysis of our model shows that it is possible to
reproduce identical results.

Let us now study the main problem we should like to solve in this work, that
is the radiation flux of the black hole and its relations with ghosts.
\section{Hawking Radiation}
In evaluating the Hawking flux we shall follow the analysis of Bilal and
Callan \cite{BC}.  We are interested in the value of the -- component of the
stress energy tensor in an asymptotically minkowskian system of coordinates.
We know that under a conformal trasformation the stress energy tensor
trasforms as:
\begin{equation}
T_{\pm\pm}(w^{\pm})=\left( \frac{\partial w^{\pm}}{\partial
x^{\pm}}\right)^{-2}
\left( T_{\pm\pm}(x^{\pm})+\frac{c_{cl}}{24}D_{x^{\pm}}^{S}(w^{\pm}) \right)
\end{equation}
\noindent
where $D_{x^{\pm}}^{S}(w^{\pm})=\;$Schwarzian derivative$\;=
\frac{{w^{\pm}}'''}{{w^{\pm}}'}-
\frac{3}{2}(\frac{{w^{\pm}}''}{{w^{\pm}}'})^{2}$ and $c_{cl}=\;$classical
central charge of the system.

We have from the costraints: $T_{\pm\pm} \equiv T_{\pm\pm}'+t_{\pm}=
T_{\pm\pm}^{\rho,\phi}+T_{\pm\pm}^{M}+t_{\pm}=0$.
\noindent
At the classical level only $T_{\pm\pm}^{\rho , \phi}$ trasform anomalously
with $c_{cl}=-12\kappa$ so we have to fix $t_{\pm}$ in order to cancel this
anomaly and obtain a total
$T_{\pm\pm}$ trasforming as a tensor and not as projective connections. This
is fundamental for the condition $T_{\pm\pm}=0$ to be a coordinate invariant
statement. Thus we have:
\begin{equation}
t_{\pm}(w^{\pm})=\left( \frac{\partial w^{\pm}}{\partial x^{\pm}}\right)^{-2}
\left( t_{\pm}(x^{\pm})+\frac{\kappa}{2} D_{\pm}^{S}(w^{\pm}) \right)
\end{equation}
In $T_{\pm\pm}^{\rho,\phi}$ only the matter generated anomalous part
${T_{\pm\pm}^{\rho,\phi}}^{I}\!=\!\kappa[(\partial_{\pm}\rho)^{2}-
\partial_{\pm}^{2}\rho]$ contributes to schwarzian anomaly and Strominger term
${T_{\pm\pm}^{\rho,\phi}}^{II}\!=\!2[\partial_{\pm}(\rho-\phi)\partial_{\pm}
(\rho-\phi)-\partial_{\pm}^{2}(\rho-\phi)]$ does not influence it. The problem
is that we use different metrics to define the measures of Polyakov term and
the Strominger one. In the first case we adopted a Weyl metric
$g_{++}\!=\!g_{--}\!=\!0$, $g_{+-}\!=\!-\frac{1}{2} e^{2\rho} \delta_{+-}$, in
the second a rescaled one $g_{++}\!=\!g_{--}\!=\!0$, $g_{+-}\!=\!-\frac{1}{2}
e^{2\rho} e^{-2\phi} \delta_{+-}$.
In our RST frame of quantization we can always (even at one loop) fix the gauge
$\rho\!=\!\phi$ so for every solution the Strominger term has a flat metric. In
calculating Hawking radiation we have to evaluate the stress energy tensor in
an asymptotically minkowskian system of coordinates. But Schwarz term vanishes
for linear trasformations between minkowskian systems.
This shows that Strominger term does not contribute.

With $t_{\pm}$ fixed as above we have:  $T_{\pm\pm}^{'}+t_{\pm}=0$
We are interested in the total value of the outgoing flux of energy on
$I_{R}^{+}$ but it is now evident that this will be equal to $t_{-}$ in
asymptotically minkowskian coordinates.

Let us now consider the dynamical solution:

\begin{equation}
\Omega=\chi=-\frac{\lambda^{2}x^{+}x^{-}}{\sqrt{\kappa-2}}-
\frac{\kappa}{4\sqrt{\kappa-2}}\ln(-\lambda^{2}x^{+}x^{-})-
\frac{m(x^{+}-x_{0}^{+})}{\lambda x_{0}^{+} \sqrt{\kappa-2}} \theta(x^{+}-
x_{0}^{+})
\end{equation}

{}From the fields transformations and the condition $\rho=\phi$ we have for
$x^{+}>x_{0}^{+}$:

\begin{equation}
\frac{e^{-2\rho}}{\sqrt{\kappa-2}}+\frac{\kappa}{2 \sqrt{\kappa-2}} \rho=
-\frac{\lambda^{2}x^{+}x^{-}}{\sqrt{\kappa-2}}-
\frac{\kappa}{4\sqrt{\kappa-2}}\ln(-\lambda^{2}x^{+}x^{-})-
\frac{m(x^{+}-x_{0}^{+})}{\lambda x_{0}^{+} \sqrt{\kappa-2}} \theta(x^{+}-
x_{0}^{+})
\end{equation}
\noindent
First we pass to an asymptotiocally minkowskian coordinate system:
\begin{equation}
\left\{\begin{array}{rcl}
e^{\lambda w^{+}}\! &=& \! \lambda x^{+}\\
e^{-\lambda w^{-}}\! &=& \! -\lambda x^{-}
\end{array}
\right.
\end{equation}
\noindent
we find from (33)(for $x^{+}>x^{+}_{0}$):
\begin{equation}
e^{-2\rho}+\frac{\kappa}{2} \rho=e^{\lambda (w^{+}-w^{-})}-\frac{\kappa
\lambda}{4}(w^{+}-w^{-})-\frac{m}{\lambda^{2} x_{0}^{+}}
\left (e^{\lambda w^{+}}-e^{w^{+}_{0}} \right)
\end{equation}
Unlike the CGHS treatment, in the present case it is clearly not
possible to find a coordinate transformation that renders the $\rho$, $\phi$
fields static (i.e. only dependent on $x^{+}\!-\!x^{-}$). What we can do is to
find a coordinate system in which $\rho$, $\phi$ are quasi-static in the $I^{+
}_{R}$ region. From the preceding equation (35) we suggest:

\begin{equation}
\left\{\begin{array}{rcl}
\tilde{w}^{+}\! &=&\! w^{+}\\
&&\\
\tilde{w}^{-} \!&=&\! -\frac{\textstyle 1}{\textstyle \lambda}
\ln \left(e^{\textstyle{-\lambda w^{-}}}\!+
\frac{\textstyle m}{\textstyle {\lambda^{2} x_{0}^{+}}} \right)
\end{array}
\right.
\end{equation}

With this coordinate transformation the leading term for
$\tilde{w}^{+}=w^{+} \rightarrow \infty$ is
$e^{\lambda(\tilde{w}^{+}-\tilde{w}^{-})}$ so $\rho$, $\phi$ are quasi static.
Then we have:

\begin{equation}
D_{w^{-}}^{S}[\tilde{w}^{-}]=-\frac{\lambda^{2}}{2} \left[ 1-\left( 1+
\frac{m}{\lambda^{2} x_{0}^{+}}e^{\lambda \tilde{w}^{-}}\right)^{2} \right]
\end{equation}

In the minkowskian system of coordinates we do not want to have any outcoming
stress energy besides the one determined by $T_{--}^{M}$ hence $t_{-}(w^{-})=0$
\footnote{Note that this statement is unambiguous only in an asymptotic
minkowskian system because transformations between this kind of system have
vanishing Schwarzian derivative.}. Finally we find:
\begin{equation}
t_{-}(\tilde{w}^{-})=\frac{1}{4} \lambda^{2}\kappa \left[ 1-\frac{\textstyle 1}
{\textstyle (1+\frac{m}{\lambda^{2} x_{0}^{+}} e^{\lambda \tilde{w}^{-}})^{2}}
\right]
\end{equation}
\noindent
This is the same result as found by CGHS \cite{CGHS}. It is also equal to the
flux found by Bilal and Callan \cite{BC} with the difference that now one has
$\kappa=\frac{N}{12}$ and so no ghost unphysical contribution.
For $\tilde{w}^{-}\! \rightarrow + \infty $ Hawking radiation is emitted at a
costant rate $\frac{1}{4} \lambda^{2} \kappa = \frac{\lambda^{2} N}{48}$.
For $\tilde{w}^{-} \!\rightarrow - \infty $ it tends to zero. Note that the
flux
is positive for every value of N.

\section{Conclusions}
This work shows that it is possible to find for the GCHS model a frame of
quantization in which one obtains a decoupling of the ghosts contribution and
a conformal invariant field theory. It is found that this model predicts
substantially the same results as the preceding works with the difference that
now the $\kappa$ value is not shifted relatively to its semiclassical value.
It is also to be noted that the solutions and the passages used are still
dependent on a shifted value $(\kappa-2=\frac{N-24}{12})$ but the results of
physical interest are not. Bilal and Callan concluding their work \cite{BC},
underlined what they have seen as the real problems of their model. The first
one was the contribution of the reparametrization ghosts to the Hawking
radiation. The second, and surely more important, was the permanently
continuing outgoing flux and thus ending up in negative mass solutions.
The aim of this work is to show that the first problem is only a matter of
opportune choice of field redefinition and, in some sense, it can be
considered as a technical detail. It is also evident that the resolution of
this point does not contribute to the solution of the second one. This model
presents, as RST and ABC do, the same asymptotically constant flux independent
of Bondi mass. To solve this puzzle, common to all dilaton gravity black hole
models, the author is at present only aware of two approaches. The first is
the RST one cited above: the solution is to choose an opportune boundary
condition for evaporating solutions as to match it to a LDV \cite{RST}.
The problem is that this brings to a kind of violation of the cosmic
censorship hypothesis and so compels us to a radical redefinition of this
assumption \cite{C.Cens}. The second, more radical approach is that of
Belgiorno, Cattaneo, Fucito, Martellini \cite{BCFM} in which the unbounded
energy problem is solved by going beyond the usual dilaton model.
Their idea is that energy unphysical result is common to all dilaton models
because it is due to a basic "bug" in the Liouville model underlying every one
of these. Those authors propose a conformal affine Toda model. So doing they
obtain a new theory that shows a realistic behaviour of physical quantities
such as energy and temperature. These results encourage the hope that 2D
quantum gravity contains a certain amount of physical information built in and
that some unphysical behaviours of actual models can be remedied.

\section*{Note}

After this article was submitted for pubblication, the author became aware of
a recent work by Strominger and Thorlacius \cite{ST}, in which a similar
conclusion is reached about the action here used (with the Strominger term).
In their generalized model, still certain amount of freedom is left by
conformal invariance, which is then fixed by the need of the coexistence of
LDV and $N$ (not $N-24$) dependent Hawking radiation.

\section*{Acknowledgements}

I wish to thank K.Yoshida for the constant help he gave me.\\
I am also grateful to M.Martellini for an illuminating discussion.

\end{document}